# Malaria incidence and prevalence: An ecological analysis through Six Sigma approach

Md. Al-Amin [1]* , Kesava Chandran Vijaya Bhaskar [2] , Walaa Enab [2] , Reza Kamali Miab [2] ,
Jennifer Slavin [2] , Nigar Sultana [2]

[1] University of Massachuesetts Lowell, Lowell, MA, USA
[2] North Carolina State University, Raleigh, NC, USA
*Corresponding Author: alamintex20@yahoo.com

## INTRODUCTION

The world started experiencing mosquitoes more than 100-98 million years ago as scientists revealed the age of the oldest fossil of a female mosquito of cretaceous period, which was discovered in Myanmar in the early 2000s [1]. The growth of human civilization due to the emergence of global commerce and urbanization brought immense changes to communication, forest and agriculture. Notably, these interconnected developments, along with rapid population growth, contributed to the worldwide dispersal of mosquito populations [2]. In this process some highly human-adapted species spread to America, Mediterranean Basin, and Africa between 15th to 18th centuries, while Asia started experiencing those during the late 19th century and in Europe those were reported to have found during mid-20th century [3]. Near about 3,000 species of mosquitoes have been found so far across the world. Among those very few have vectorial capacity to spread those into humans and animals [4]. Yet these vectors are responsible for many deadly human diseases, such as malaria, filariasis, dengue, yellow fever, and West Nile [5]. Among these, malaria is the most fatal one that dominated the world with several subsequent epidemics including in Asia, Europe, Africa, and America since 18th century [6]. The archeological evidence showed the existence of malaria parasites even 10,000 years ago, whereas historical evidence suggests its dominance in Roman empire period and then it shook the world with several epidemics across the world before the modern era [7]. A species of parasites *plasmodium* cause malaria to humans, which are transmitted by *anopheles* mosquitoes. There are six species of *anopheles* mosquitoes, such as *plasmodium falciparum, plasmodium vivax, plasmodium ovale, plasmodium malariae*, *plasmodium knowlesi*, and *plasmodium cynomolgi* [8]. According to the world malaria report 2022 of World Health Organization (WHO), about 247 million malaria cases have been reported in 2021, which caused 619,000 fatalities worldwide, most of those occurred in sub-Saharan Africa. Approximately half of the world's population is susceptible to malaria, especially young children and pregnant women [9]. It was reported that age is a significant factor in malaria infection and children of six-10 years of age are mostly at risk [10]. It was also found that having malaria at these ages can potentially lead to some other physical malfunctions, such as anemia [10]. However, children under five years of age are also vulnerable to malaria as they are yet to grow their full immunity against the malaria parasites [11]. There are several environmental factors that are associated with malaria infection, such as rainfall, temperature, and vegetation. Usually, higher rainfall, lower temperature, increased vegetation, and residing more than five minutes away from health facility are associated with increased risk of malaria [12]. Use of malaria prophylaxis was found to be a crucial factor associated with lower risk of malaria infection while mothers' education level and employment status were also associated with malaria risk [12]. Moreover, travelers visiting to the highly malaria endemic areas have higher risk of malaria infection as they have less knowledge about malaria [13].

Malaria has a deeper impact on personal and social lives. Pregnant women having suffered from intra-uterine growth retardation due to malaria have risk of abortion and giving birth to children with low weight causing increased neonatal mortality. Malaria survivors experience behavioral and cognitive disorders, measles, malnutrition, and respiratory diseases in the long run [14]. Apart from physical sufferings, it has socio-economic impact, too [15, 16]. In a recent study, it was found that on an average diagnosis of malaria cost $6.06 and treatments cost $9.31 to $89.93 per person [17]. Center for Disease Control and Prevention estimated the direct cost of malaria was $12 billion every year [18]. WHO is continuously putting its effort to control and eradicate malaria through the global malaria program that sets an aim to eradicate malaria from the world by 2030 with its long-term visionary initiative "Global technical strategy for malaria 2016–2030" [9]. Additionally, there are several organizations that have been funding to prevent and control malaria worldwide, such as the US Agency for International Development, World Bank, and Bill and Melinda Gates Foundation [19-23].

Malaria surveillance and control measures must be the regular and continuous processes as there is a strong chance of malaria epidemics to bounce back even after the complete eradication [24]. However, WHO updated the integrated vector management (IVM) in their recent report of 2023, that recommended insecticide treated nets (ITNs), long-lasting insecticidal nets (LLINs), indoor residual spraying (IRS), larviciding, and house screening as the major preventive measures to control malaria vectors. Among those ITNs and LLINs were strongly recommended [9, 25]. Several research have been made on the efficacy of ITNs and other preventive and control measures in the recent past based on small local sample quantity and review of previous studies [26-31]. Additionally, very few research works did comparative studies on major preventive and control measures based on meta-analysis [32, 33]. No to very few studies did the holistic approach to investigate the comparative efficacy of preventive and control measures against malaria. Moreover, a holistic approach using global data resources to quantify the effectiveness of contemporary practicing measures for malaria prevention and control is very limited. Additionally, even amid the ongoing global crisis due to COVID-19, the prevalence of malaria has also increased over the last couple of years (2020 and 2021) across the malaria prone areas [9], which justifies the need for the quantification on the effectiveness of available preventive and control measures to provide the people with best preventive solutions. Therefore, this study purposes to investigate the preventive and control measures against malaria through a proven scientific approach using global data resources collected by the respective stakeholders. This study intends to use Six Sigma approach, which is a proven high performance and data driven approach that identifies and analyzes the root-causes of problems in order to solve those problems [34]. Pertaining to the nature of variables and purpose of the research, this study involves DMAIC (define, measure, analysis, improvement, and control) process of Six Sigma. DMAIC is a closed-loop framework that involves regression analysis and helps identify and eliminate the unproductive variations and look for the continuous improvement and control [35].

## METHODOLOGY

Six Sigma was introduced to reduce the defects by statistical analyses, especially for manufacturing industry, however, in contemporary research paradigm it has been proven to be an effective tool across different disciplines to identify cause-effect of problem, analysis of problems, and to suggest the realistic solution to those [36]. DMAIC process of Six Sigma guides this research study encompassing the entire methodology, which is based on five stages, such as define, measurement, analysis, improvement, and control. The define stage explores the variables related to malaria prevention and control using different tools, such as fishbone diagram, which is also known as cause effect diagram. Measurement stage explains the data collection and analysis procedures. Analysis stage covers the data analysis process. Improvement stage covers the discussion and recommendations for the improvement. Finally, control stage discusses how to sustain the goal for the eradication of mortality rate due to malaria.

### Define Stage

Malaria is dominated by several factors as stated previously. Six Sigma uses cause effect or fishbone diagram to explore those variables. Fishbone diagram identifies the root causes based on four "M", those are "method", "machine", "material", and "measurement". However, this study also includes "environment" considering it as a fundamental aspect of malaria prevention and control research. Based on the fishbone diagram (**Figure 1**), previous malaria intervention survey and literature study following factors have been found to be associated with malaria intervention.

The cause effect diagram or fishbone diagram shows the cause and effect of malaria in a single frame. Based on this cause effect diagram and findings from malaria survey analysis, following terms/factors are found to be associated with malaria prevalence and interventions.

### *Mosquito nets*

WHO highly recommends using nets while sleeping, especially among the population in high-risk areas for malaria transmission. Apart from conventional untreated mosquito nets, there are two types of nets recommended by WHO-ITNs and LLINs [37].

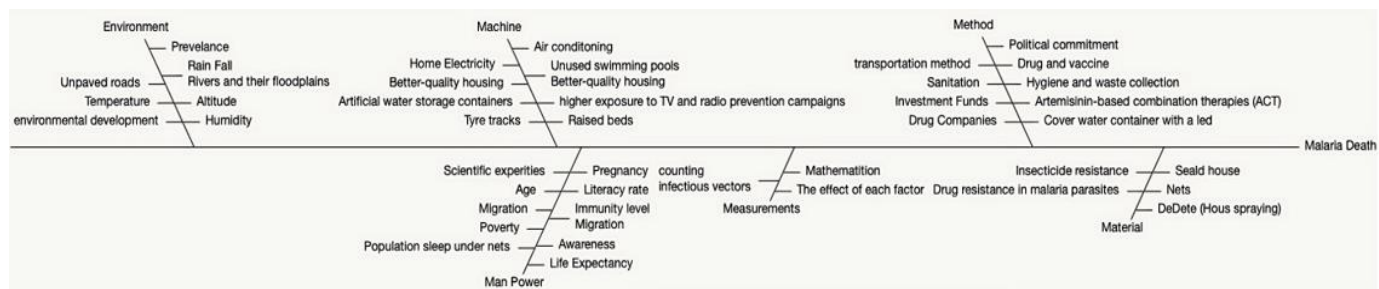

**Figure 1.** Fishbone/cause-effect diagram of malaria (Source: Authors' own elaboration)

*Insecticide treated nets*

Textile nets are coated with insecticides to prevent mosquito without compromising mechanical properties known as ITNs [38-40]. An ITN is an important tool to prevent malaria, which offers both physical and chemical barriers against mosquitoes [41]. Several trials found ITNs effective to reduce *plasmodium falciparum* prevalence among children under five years [42]. The use of ITNs has been effective in reducing malaria mortality (17.0%) and morbidity (43.0%) in children under five years and provides protection to pregnant women susceptible to malaria [43]. In general, ITNs are found to be 68.0% effective to control malaria across Africa [44].

*Long-lasting insecticidal nets*

The use of LLIN including insecticide-treated hammock net is one of the front-line interventions of global malaria prevention. Although few studies have found LLINs significantly prevent malaria, further study is critical to find out the comparative efficacy of this net [37].

*Populations having access to nets & insecticide treated nets*

WHO recommends that in order to significantly reduce the occurrence of malaria that 3.2 billion people need to sleep under mosquito nets daily in areas at high risk of contracting malaria. This has made sleeping under an ITN a crucial intervention factor. Moreover, the latest surveys showed that the access of population to ITNs varies from 6.9% (Dijibouti) to 83.1% (Mali) averaging 51.3% across Africa [45].

*Age*

Transmission intensity is associated with the average age at which children experience serious incidents of malaria. Children suffer more from symptomatic malaria at a younger age in areas of high transmission [19]. Usually, the risk of malaria incidence is positively related to the age, meaning people have more chance of getting malaria incidence the older they are [41].

*Malaria prevalence*

Malaria prevalence is one of the major factors in surveying the malaria incidence. It signifies the infestation of malaria infected mosquitoes in certain areas and its ultimate consequences including mortality and morbidity rate [47].

*Children under five years sleeping under nets & insecticide treated nets*

The mortality and morbidity rate due to malaria among the children under five years has been found to be very prevalent in severe malaria transmission areas.

WHO recommends all children under five to use ITNs, particularly LLINs, which are cost-effective. It is estimated that this alone can reduce malaria prevalence among children under five years by approximately 50.0% and all-causes of mortality by 17.0% [48, 49].

*Percentage using indoor residual spraying*

According to WHO, IRS can contribute to eliminating malaria if rigorously applied. As a vector control measure, the percentage of population using IRS is another key factor in IVM [50].

*Education*

Education has a positive impact on the prevention of malaria in sub-Saharan Africa Region. Educated parents are found to be more aware of the prevention and treatment against malaria for their [12].

*Sanitation facility*

Sanitation is a well-known factor to prevent the infestation of any kind of mosquito [51].

*Life expectancy*

Low life expectancy at birth is the likelihood of a child to have problems with health care and reduces years that a child is expected to live. High life expectancy at birth tends to have minor problems with health care and longevity [52]. On the contrary, malaria decreases the life expectancy at birth by six years in the malaria prevalent regions [53].

**Measurement Stage**

Measurement stage complies with data collection procedures of conventional research method. Moreover, it also involves the data analysis procedure plans and measurement system analysis (MSA). However, MSA was excluded owing to its limitations in reproducibility and repeatability benefits.

**Data Collection Plan**

Data were collected from several leading stakeholders in malaria survey and intervention, such as WHO [54], UNICEF [55], ourworldindata.com [56], and statcompiler.com [57]. As Sub-Saharan African region is the biggest sufferer of malaria [58], originally, data for all African countries were compiled, along with each country's population and populations of children under five. The occurrence of malaria for each country was then added to the dataset along with the number of deaths from malaria. Statcompiler.com was then used to see what data were available along with the factors for each country. Corresponding population, malaria cases, and malaria deaths were added. The percentages were initially analyzed for linear regression. However little to no correlation was found. Looking at the data again the authors decided to convert to a per capita number and actual number of people in each category to see if a correlation could be determined. Then other factors such as literacy and artemisinin-based combination therapy were added to see if there was a correlation or interdependence. Since the data came from multiple sources, was collected by hundreds of individuals, was acquired through surveys, and there was no clear method of collecting data, hence, there was greater chance of inconsistent data accuracy. Therefore, conducting MSA, which is used to see if the data have reproducibility and repeatability would not be beneficial for this study.

**Table 1.** Summary of fit, ANOVA, & parameter estimate for linear regression

| Variables/factors | $R^2$ | Root mean square error | t-value | p-value |
|---|---|---|---|---|
| Malaria prevalence | 0.977125 | 6367.075 | 26.95 | <0.0001 |
| Net usage | 0.908886 | 12707.400 | 13.02 | <0.0001 |
| IRS usage | 0.025509 | 41557.750 | 0.67 | 0.5137 |
| Literate population | 0.637569 | 25344.020 | 5.47 | <0.0001 |
| Unimproved sanitation | 0.675367 | 23986.070 | 5.95 | <0.0001 |
| Life expectancy | 0.301148 | 35192.940 | 2.71 | <0.0150 |

### Analysis Stage

Analysis stage involves the selection of dependent and independent variables, statistical analysis procedures, and model formulations.

### Variables/Factors

Since the malaria deaths in the entire population were recorded and analyzed, the response variable was the number of deaths in the associated year of survey data collected for each country. The independent variables considered in this analysis include several socio-economic and demographic factors. The demographic factors associated with the malaria deaths in the population was the life expectancy in each country. The socio-economic variables were malaria prevalence in children, usage of net, usage of IRS, literate population, population with unimproved sanitation facility. The sub factors associated with net usage of nets and malaria prevalence in children also comes under this category.

### Statistical Analysis

The present analysis aimed to establish a linear relationship between the dependent and the independent variables. The analysis of this model was done using JMP 16 Pro software (SAS, NC, USA). JMP provides four types of analysis for linear regression, such as summary of fit, parameter estimates, analysis of variance (ANOVA), and residual plots. Summary of fit is mostly described by R-squared value, whereas parameter estimate, and ANOVA are described by p-value and F-value. Since this is a linear regression model, the model F-value and the p-value will be the same. The values are calculated based on the hypothesis testing of the slope. If the p-value and the F-value is found to be less than 0.05 (5.0%), then those factors are considered to be significant. The significant factors qualify for multiple linear regression in the next stage. The p-value of the hypothesis test to check whether the independent variables are significantly associated with the malaria deaths or not. The variables from the linear regression with a p-value of less than 5% level of significance were included in the multiple regression analysis. Sub factors for the factors that were identified significant from the multiple regression analysis were identified. With these set of subfactors again linear regression was performed to identify the statistically significant sub-factors. The sub-factors were included in the multiple regression analysis.

### Model Formulations

The model for linear regression and multiple linear regression follows Eq. (1) and Eq. (2), respectively. The model is built between the dependent and the independent variable. Hypothesis testing is carried out for the slope of the model coefficient.

$$Y = \beta_0 + \beta_1 x + \varepsilon \quad (1)$$
$$Y = \beta_0 + \beta_1 x_1 + \cdots + \beta_n x_n + \varepsilon \quad (2)$$

where $\beta_0$ is y-intercept, $\beta_1$ is slope of the line, and $\varepsilon$ is error variable.

The p-value is used to test the hypothesis that there is no relationship between the predictor and the response. Or, stated differently, the p-value is used to test the hypothesis that true slope coefficient is zero. Statistically significant factors are identified by checking the p-value. If the p-value is found to be less than 5%, then those factors are considered as statistically significant. The mean of the errors should be equal to zero. The probability plots of the residuals should be normal. If the standard deviation of the residuals is less, then the data fits the model very well. The normality of the residuals is also checked for validation and assessing the multiple linear regression model.

## ANALYSES OF RESULTS

### Linear Regression

At first, linear regression was conducted. The linear regression includes the selected factors from fishbone diagram and data sources, such as malaria prevalence, net usage, IRS usage, literate population, unimproved sanitation, and life expectancy.

**Table 1** shows the corresponding R-squared value, which is the percentage of variation in malaria death explained by the corresponding factor. This study maintained a minimum of 50% coefficient of variation requirement in the model. Thus, the variables IRS usage and life expectancy do not possess the required R-squared value and will not be considered for further analysis. Moreover, significant factors are selected for multiple linear regression. Thus, from p-value and F-value, factors such as malaria prevalence, net usage, literate population, unimproved sanitation, and life expectancy were found to be significant (<.050). However, due to the limitation on the R-squared value, the factor 'life expectancy' is eliminated. The one outlier that appears in most the charts is Nigeria. It has higher population and higher population density than all the other countries in Africa. For this data point there is probably an additional factor that is influencing this datapoint.

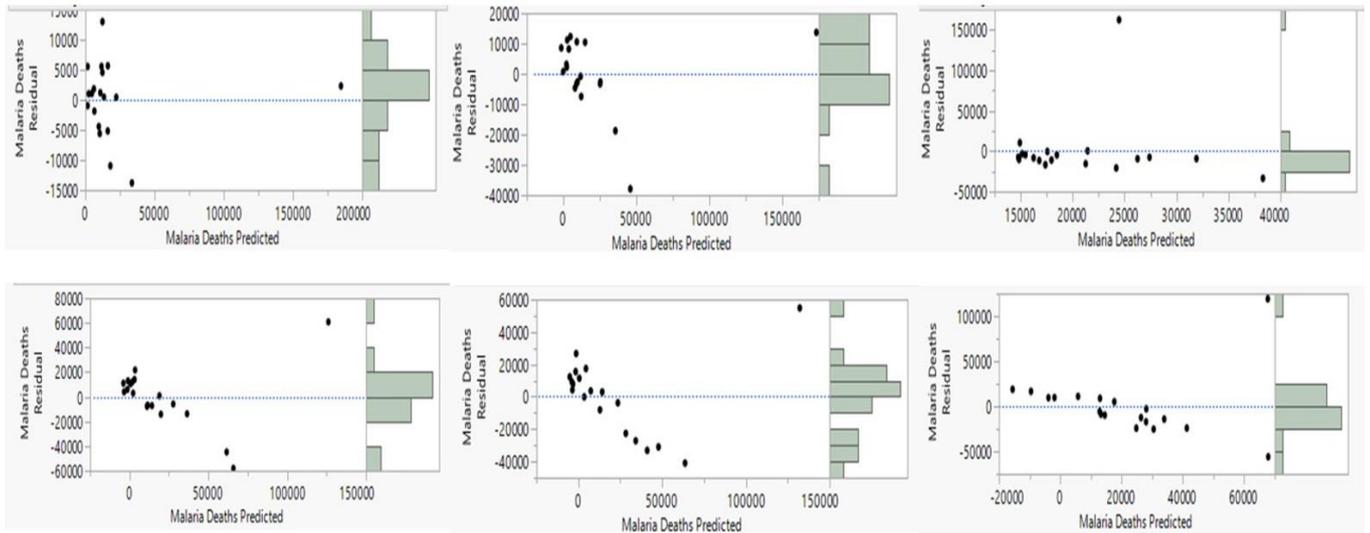

**Figure 2.** Residual & pareto plots (predicted & by X-plot) in linear regression (Source: Authors' own elaboration, using JMP 16 Pro statistical analysis software)

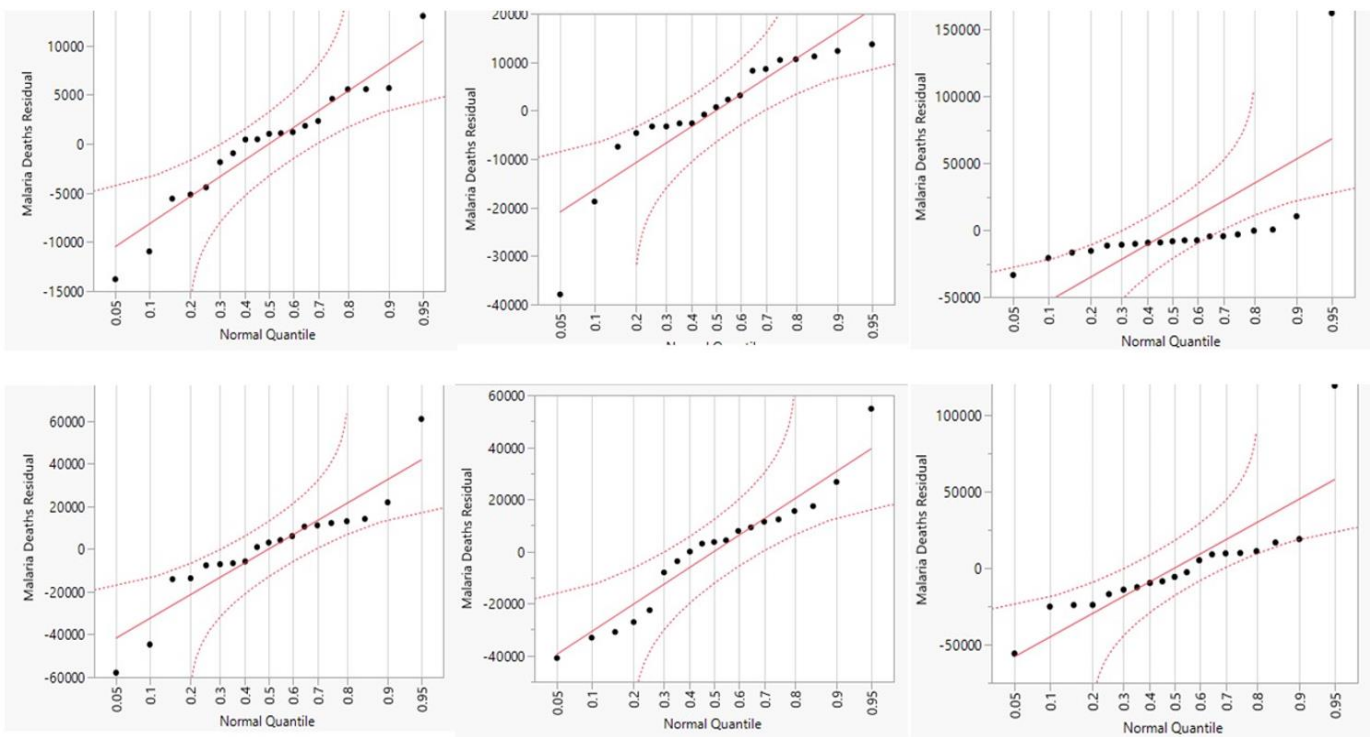

**Figure 3.** Bivariate fit & residual normal quantile plot in linear regression (Source: Authors' own elaboration, using JMP 16 Pro statistical analysis software)

The calculated residuals should be normally distributed having a mean of zero is one of the steps in testing the significance of the factors. The standard deviation of the residuals of a particular factor if minimum, then the corresponding factor is found to be significant.

In JMP results (**Figure 2** and **Figure 3**), the residual by predicted plot, actual by predicted plot, residual by row plot, residual by X plot and residual normal quantile plot are the diagnostic plots for the residuals. The two main plots of focus are the residual by predicted plot and the normal quantile plot (**Figure 3**). The former plot is used to provide graphical explanation of the residual mean and the latter is used to provide a graphical explanation of the normality of the residuals. If the plotted points have a spread closer to the linear line and in between the confidence intervals, then the corresponding factor has its residuals following a normal distribution. Thus, based on summary of fit, parameter estimate hypothesis testing and residual diagnostics, it has been found that malaria prevalence, net usage, literate

**Table 2.** Parameter estimates, effect tests, & effect summary

| Variables/factors | Log worth | p-value |
|---|---|---|
| Malaria prevalence | 3.042 | 0.00091 |
| Net usage | 1.669 | 0.02144 |
| Literate population | 0.705 | 0.19718 |
| Unimproved sanitation | 0.094 | 0.80547 |

population, and unimproved sanitation are found to be significant.

**Multiple Linear Regression of Malaria Death vs. Major Factors**

ANOVA assessed the entire model. The entire model is categorized as significant if the F-value is found to be less than 0.05. ANOVA finds the F-value<0.0001 and thus the model is significant. **Table 2** shows the statistical significance of the individual parameter. As the log worth decreases the p-value and the corresponding F-value increases. Thus, **Table 2** concludes that the factors malaria prevalence among children under age five years and the net usage are significant. The pareto plot of the transformed estimates ranks the factors by their orthogonal estimate (**Figure 3**). On comparison of the deaths with respect to age, the children under the age of five were highest contributors to the malaria deaths in the African countries. The underlying factors of children under the age of five and the net usage are used as sub factors for further analysis. Therefore, moving forward for multiple linear regression, the dependent factors were selected to be access to nets, ITN net usage, LLIN net usage, children slept under any net, children slept under ITN, children slept under LLIN while the independent factor is malaria death.

**Malaria Death vs. Sub-Factors**

The procedure of analysis is as same as mentioned in linear regression. From **Table 3**, shows the factors access to nets, ITN net usage, children who slept under any net, children slept under ITN were found to be significant. The hypothesis of the slope is tested, and the results are mentioned in **Table 3**. Thus, taking reference from linear regression results. and **Table 3**, access to nets, ITN net usage, children slept under any net, and children slept under ITN are the significant factors.

**Table 3.** Summary of fit & ANOVA

| Variables/factors | $R^2$ | Adjusted $R^2$ | p-value |
|---|---|---|---|
| Access to nets | 0.852647 | 0.843979 | <0.0001 |
| ITN net usage | 0.863194 | 0.855146 | <0.0001 |
| LLIN net usage | 0.028101 | -0.029070 | 0.4927 |
| Children slept under any net | 0.886965 | 0.880316 | <0.0001 |
| Children slept under ITN | 0.903170 | 0.897474 | <0.0001 |
| Children slept under LLIN | 0.040006 | -0.01646 | 0.4116 |

The mean of the residuals, standard deviation, and the normality of the residuals are checked (**Figure 4** and **Figure 5**) from the normal quantile plot. The factors that were found significant in linear regression were proved to be significant in this section too. Thus, using the significant factors, the multiple regression analysis was done.

**Malaria Death vs. Significant Sub-Factors**

The R-squared value was found to be 0.969549 and the adjusted R-squared value was found to be 0.960848. The difference between R-squared and the adjusted R-squared was found to be minimum and thus it is concluded that right factors were chosen to build the model (**Table 3**). F-value mentioned in JMP results was found to be <0.0001. This states that the model as a whole is significant.

**Table 4** shows the factors ITN net usage (p=.010<.050) and children slept under ITN (p=.005<.050) were statistically significant.

**Interpretation of Prediction Equation**

**Table 4** shows the estimate values of the significant factors ITN net usage and children slept under ITN are -0.007731 and 0.058770, which interpret, as follows.

- For every 1,000 children who sleep under an ITN net, the malaria death increased by approximately 59.
- For each 1,000 people using ITN, the malaria deaths decreased by approximately eight units. The results are consistent with the recently published literature [49, 59, 60].

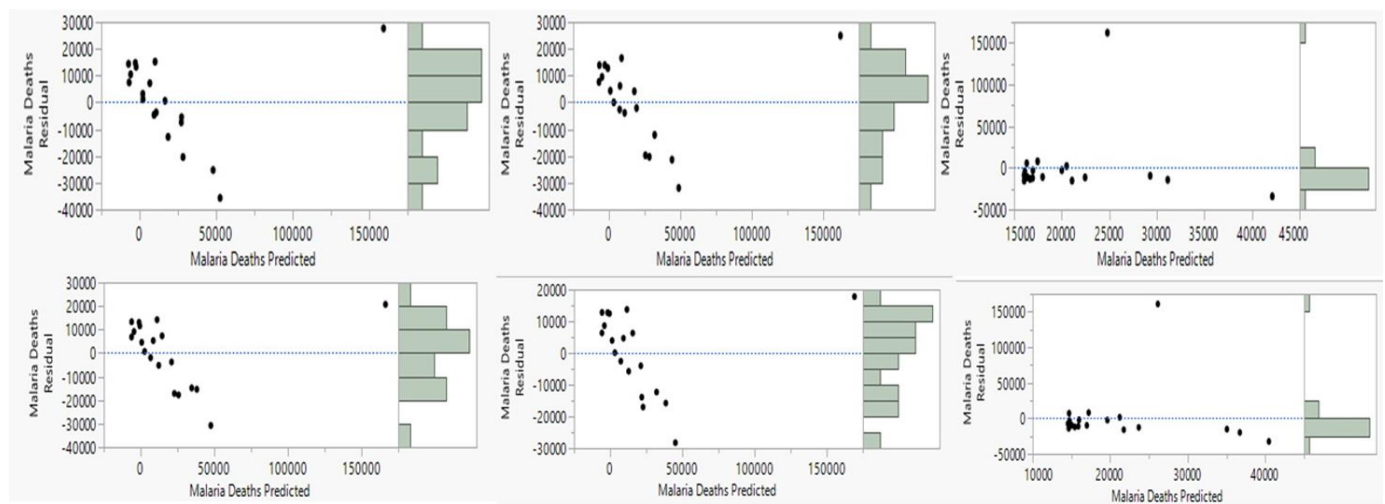

**Figure 4.** Residual & pareto plots (predicted & by X-plot) in multiple linear regression (Source: Authors' own elaboration, using JMP 16 Pro statistical analysis software)

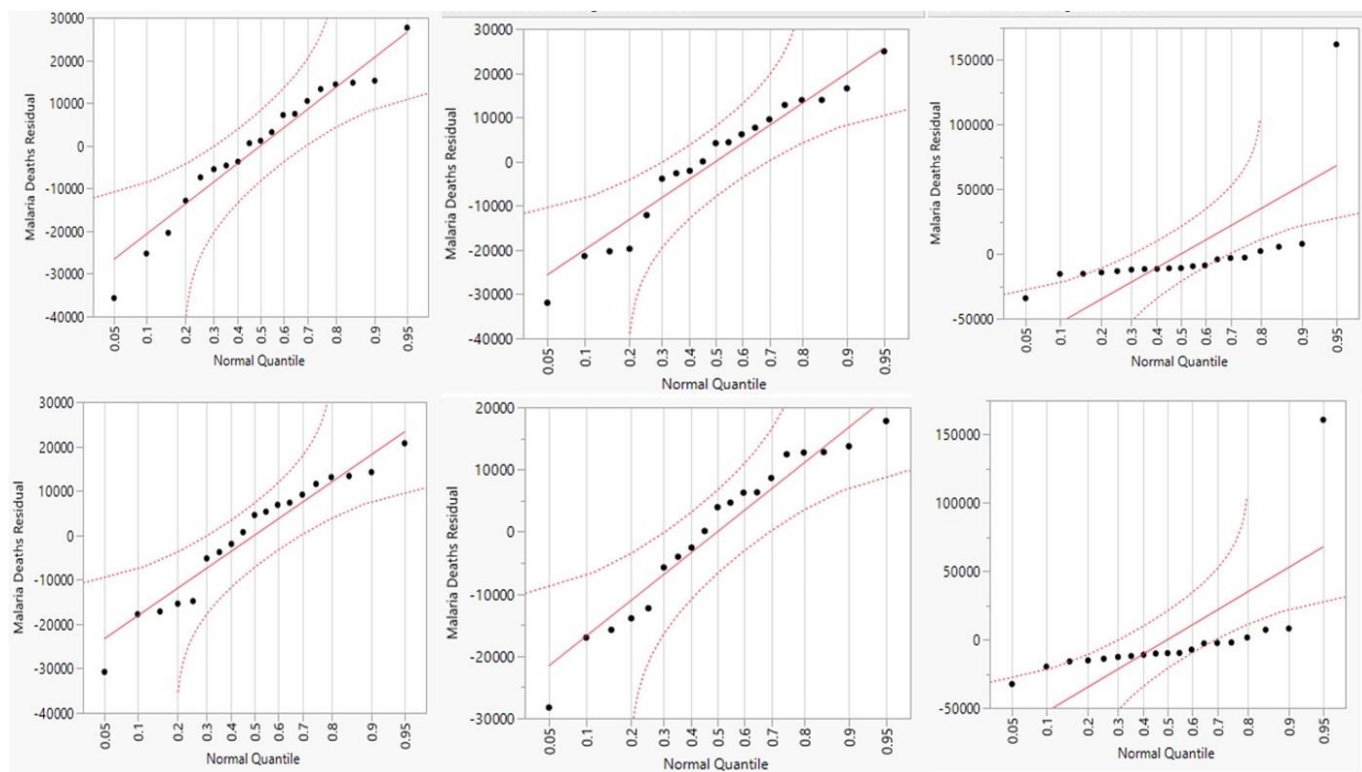

**Figure 5.** Bivariate fit & residual normal quantile plot in multiple linear regression (Source: Authors' own elaboration, using JMP 16 Pro statistical analysis software)

**Table 4.** Parameter estimates, effect tests, & effect summary

| Variables/factors | Estimate | Log worth | p-value |
|---|---|---|---|
| Access to nets | 0.001376 | 0.623 | 0.23825 |
| ITN net usage | -0.007731 | 1.940 | 0.01147 |
| Children slept under any net | -0.017841 | 0.495 | 0.31998 |
| Children slept under ITN | 0.058770 | 3.231 | 0.00059 |

The correlation of population using ITN with malaria death is theoretically correct as it has negative sign. However, the correlation of children who slept under ITN is positive, which is incorrect. This might be due to incorrect data collection, or the fact that parents may have provided incorrect responses under assumptions that they could be penalized for not making children sleep under nets.

**Improvement Stage**

The following recommendations are made by the authors based on this research and subsequent analysis for further improvements in malaria prevention and control initiatives.

- Benchmarking the countries that have had a significant reduction in malaria cases and see if their methods could be implemented in other countries.
- Identifying countries that have a high occurrence of malaria and low usage of nets and prioritize net distribution to those countries.
- Analyzing the areas, where population density is high, but the usage of nets is low. Prioritize the mass distribution of nets to these areas. The higher population density could make net distribution easier and have a greater effect than in sparsely populated areas.
- Creating additional database fields for brands, condition, and age of nets in surveys.
- As surveys are conducted with the households, distribute nets at the same time to those households who are at high risk and do not have any-especially those with children under five.
- If nets were mass distributed two-three years ago, it is probably time to redistribute new nets. Keeping a database of when nets were distributed.
- Obtaining feedback from the people. If people are not using nets, ask why and then develop plans to correct for example, if they are not using nets because they are difficult to use, develop an easier to use net. If they have difficulty persuading their children to sleep under nets, create fun patterns that children like.
- Continuing educating the population on the benefits of using nets.

**Control Stage**

The recommended control methods are, as follows.

- Continuing to monitor, analyze, and report on the existing indicators.
- Creating a database for brands of nets, so that it can easily be cross referenced after surveys are conducted to know if the net is an ITN.
- Monitoring malaria rates in low occurrence countries especially those that border on a country with high occurrence and make sure their numbers are not rising.

### Future Research Directions & Limitations

During this research study it became clear that better methods of data mining and collection are needed. Creating a database that holds the number of malaria cases and deaths, the country's population and the usage of interventions all in one place by year would be beneficial for future study. Other potential thoughts for continuing the prevention and control of malaria are, as follows:

- effectiveness of the vaccine,
- effectiveness of the type of net,
- determining if a more effective/longer lasting net can be manufactured,
- investigating why children sleeping under nets had a positive correction to malaria deaths, and
- investigating what other factors are causing Nigeria to be an outlier from the other African countries.

The reason why data on every country for each year was unavailable was because not all of the chosen countries conducted surveys on their populations every year. Furthermore, it took time for organizations to compile and publish the data, resulting in varying time frames for when surveys were conducted. This made it challenging and time-consuming to organize all of the data points. Eventually, the authors decided to group the data by years, using surveys completed between 2017-2020 as the most recent example. However, not all countries had reported data for this time frame, so data from as far back as 2001 was collected for some countries. During the data mining process, the authors realized that a more precise correlation could be achieved by comparing the data to the population of the year it was collected. They identified 20 countries with current data (2017-2020), and although Nigeria was identified as an outlier due to having comparatively severe impact of malaria than other sub-Saharan African countries, the authors did not exclude it from the list of sample countries during statistical analysis. Moreover, the statistics on children who slept under nets was incorrect assuming the parents who were surveyed did not report the data appropriately due to political or social limitations.

### Policy Implications

The outcomes derived from this ecological analysis offer valuable insights for shaping global malaria prevention and control policies. The evident advantages of ITNs highlighted in this research endorse the ongoing endeavors to enhance the accessibility and utilization of ITNs on a broader scale, particularly targeting vulnerable populations, such as children under five and pregnant women. Nations facing low ITN coverage are encouraged to prioritize the distribution of nets and implement behavior change communication initiatives to promote increased net usage. Furthermore, the sustained efficacy of nets, even in the presence of some insecticide resistance, underscores the importance of retaining ITNs as a pivotal intervention, while carefully monitoring any potential decline in effectiveness. In conjunction with ITNs, additional measures like IRS, sanitation enhancements, and educational initiatives are advantageous and should be integrated as complementary strategies. The design of IVM approaches should be tailored to local resistance patterns, guided by continuous surveillance efforts. Crucially, maintaining the progress achieved in alleviating the global burden of malaria requires unwavering political commitment and sustained financial support.

## CONCLUSIONS

Malaria is a preventable and treatable disease. However, it remains a major public health problem in many parts of the world, especially in sub-Saharan Africa. In 2020, there were an estimated 241 million cases of malaria and 627,000 deaths as WHO reported. The preventive and control measures of malaria have proven to be effective in reducing the transmission and burden of this deadly disease including regular nets, ITNs, LLINs, and IRS. Moreover, improved sanitation and literacy also play a crucial role in malaria prevention and control. The data analysis by Six Sigma DMAIC process of most malaria prevalent countries of sub-Saharan African Region obtained by world recognized organizations showed ITNs was the most effective preventive and control measure among all factors as use of ITNs can decrease malaria mortality up to eight out of 1,000 people. It suggests more rigorous research on the versatile application of ITNs globally. Apart from statistical significance, this study introduces the Six Sigma process in malaria research, which might be a leading role in future malaria research for the respective researchers and stake holders. However, inconsistency in terms of data and survey period is a limitation of this study, which needs to be overcome in future research. Despite the progress made in malaria control, the disease continues to pose a significant public health challenge in many parts of the world. There is a need for continued investment in research, surveillance, and innovative strategies to sustain and expand the gains achieved in malaria prevention and control.

**Author contributions:** All co-authors have been involved in all stages of this study while preparing the final version. They all agree with the results and conclusions.

**Funding:** No funding source is reported for this study.

**Declaration of interest:** No conflict of interest is declared by the authors.

**Ethical statement:** The authors stated that the study did not require any ethical approval. No live subjects were involved in the research.

**Data sharing statement:** Data supporting the findings and conclusions are available upon request from corresponding author.